\documentclass[amsmath,amssymb,twocolumn,superscriptaddress]{revtex4}

\usepackage{graphicx}
\usepackage{dcolumn}

\def\be{\begin{equation}}
\def\ee{\end{equation}}
\def\beq{\begin{eqnarray}}
\def\eeq{\end{eqnarray}}

\usepackage{bm}
\usepackage{graphicx}
\usepackage{dcolumn}
\usepackage{amsmath,enumitem}
\usepackage[utf8]{inputenc}
\usepackage{graphicx, psfrag}
\usepackage{amssymb}
\usepackage[colorlinks=true, citecolor=blue, urlcolor = blue, linkcolor= red, bookmarks=true]{hyperref}
\usepackage{float}
\usepackage{amsmath}
\usepackage{tensor}
\usepackage{amsfonts}
\usepackage{dcolumn}
\usepackage{hyperref}
\usepackage{subfigure}
\usepackage{pgfplots}
\usepackage{epstopdf}
\usepackage{booktabs}
\usepackage{amsmath, amssymb}
\usepackage{multirow}
\usepackage{graphicx}
\usepackage[usenames,dvipsnames]{xcolor}
\newcommand{\rev}[1]{\textcolor{blue}{#1}}

\begin{document}
\title{Transient Dynamical Wormholes with Decaying Radial Energy Flux}

\author{Allah Ditta}
\email{mradshahid01@gmail.com}
\affiliation{School of Science, Walailak University, Nakhon Si Thammarat, 80160, Thailand}
\affiliation{Center of Excellence in High Energy Physics, Faculty of Science, Chulalongkorn University, Phayathai Road, Pathumwan, Bangkok 10330, Thailand}

\author{Phongpichit Channuie}
\email{phongpichit.ch@mail.wu.ac.th(\textcolor{red}{Corresponding Author})}
\affiliation{School of Science, Walailak University, Nakhon Si Thammarat, 80160, Thailand}
\affiliation{College of Graduate Studies, Walailak University, Nakhon Si Thammarat, 80160, Thailand}

\begin{abstract}
We investigate a class of time-dependent traversable wormholes within the framework of general relativity by allowing the shape function to vary with time. In this setting, the evolution of the geometry is directly connected to a radial energy flux through the off-diagonal component of the Einstein field equations, providing a natural mechanism for non-static configurations. By adopting a decaying flux profile, we obtain exact solutions in which the geometry consists of a static background supplemented by a transient term that diminishes with time. The resulting spacetime satisfies the standard conditions required for a traversable wormhole, including the presence of a throat, the flaring-out condition, and asymptotic flatness. The corresponding matter content is determined explicitly and shows a clear temporal evolution governed by the decay parameter. An analysis of the energy conditions indicates that the null energy condition is violated in the vicinity of the throat, although the degree of violation decreases as the system evolves. This feature is examined further through a volume integral that measures the total amount of exotic matter, demonstrating that it approaches a constant value at late times. We also study the response of the system to small perturbations and find that, for a suitable choice of parameters, the configuration remains stable with perturbations decaying over time. We discuss how this flux-driven mechanism relates to the alternative and extensively studied approach in which wormhole evolution is instead carried by a cosmological scale factor, and we identify possible physical origins of the assumed radial flux in terms of null-fluid matter sources. Overall, the model describes a wormhole spacetime whose evolution is controlled by energy transport, leading to a gradual transition toward a static configuration. This framework offers a simple and physically motivated approach to dynamical wormholes and may be useful for exploring more realistic scenarios in both general relativity and extended theories of gravity.
\end{abstract}

\date{\today}

\maketitle

\section{Introduction}

Wormholes are hypothetical geometric structures that connect two distinct regions of spacetime and arise as solutions of the Einstein field equations \cite{EinsteinRosen1935}. Early realizations of traversable wormhole-like geometries were explored in scalar field models, where it was shown that horizonless configurations can exist under suitable conditions \cite{Ellis1973,Bronnikov1973}. A major development in the field was achieved by Morris and Thorne \cite{MorrisThorne1988}, who formulated the conditions for traversable wormholes in static and spherically symmetric spacetimes. Their analysis demonstrated that the existence of a wormhole throat requires the violation of the null energy condition (NEC), implying the presence of exotic matter \cite{MorrisThorneYurtsever1988,Visser1996}. This requirement has been extensively studied and remains a central issue in wormhole physics.

Most wormhole solutions in general relativity are based on static configurations supported by anisotropic matter distributions, where the geometry is prescribed through an appropriate choice of the shape function \cite{Visser1989}. In addition, modified theories of gravity have been widely investigated as alternative frameworks in which wormhole solutions may alleviate or reinterpret the violation of energy conditions \cite{LoboOliveira2009,HarkoLoboMakSushkov2013}. In recent years, wormhole research has expanded significantly beyond static configurations, with increasing attention devoted to dynamical solutions and their physical interpretation \cite{Hayward2009, HochbergVisser1998, Hayward1999, Roman1993}. In particular, evolving wormhole geometries have been investigated in both general relativity and thin-shell constructions \cite{GarciaLoboVisser2012}. In parallel, several studies have explored the possibility of reducing or reinterpreting the violation of energy conditions through modified gravity theories \cite{LoboOliveira2009,HarkoLoboMakSushkov2013}. Despite these advances, most existing models introduce time dependence at the level of the metric functions without identifying a clear physical mechanism responsible for the evolution.

Despite these developments, realistic gravitational systems are inherently dynamical, and it is therefore important to explore time-dependent wormhole configurations. Early investigations of evolving wormholes were presented in \cite{Kar1994}. More general dynamical wormhole geometries, including those with explicit dependence on both radial and temporal coordinates, have been studied in several works \cite{HochbergVisser1998,Hayward1999,Hayward2009,ArellanoLobo2006, GarciaLoboVisser2012}. These studies highlight the role of time dependence in determining the physical properties and stability of wormhole spacetimes. Although these studies provide valuable insights into the behavior of time-dependent wormhole geometries, the dynamical evolution is typically introduced through a prescribed metric ansatz or through specific matter sources. In particular, the time dependence is often imposed at the level of the geometry, and the corresponding matter content is subsequently determined from the field equations. As a result, the physical mechanism responsible for the evolution is not always explicitly identified. In contrast, the present work adopts a different perspective by relating the temporal variation of the geometry directly to a radial energy flux through the off-diagonal field equation. In this way, the evolution of the wormhole is not imposed a priori, but emerges naturally from an underlying energy transport process. This provides a more direct physical interpretation of the dynamical behavior and allows the geometry, matter distribution, and flux to be treated in a unified framework.

In most existing approaches, the time dependence is introduced at the level of the metric functions without a clear physical mechanism responsible for the evolution. However, within general relativity, any time variation of the geometry must be accompanied by a corresponding flow of energy-momentum. In particular, the off-diagonal component of the Einstein field equations naturally relates the time derivative of the shape function to a radial energy flux, indicating that energy transport plays a fundamental role in dynamical wormhole spacetimes \cite{HochbergVisser1998}.

A separate and long-standing line of research treats wormhole evolution through a cosmological scale factor $a(t)$ multiplying the spatial sector of the metric, so that the geometry inherits its time dependence from the expansion of a cosmological background rather than from a flux term appearing directly in the field equations. Kar and Sahdev \cite{KarSahdev1996} constructed evolving Lorentzian wormholes of this type for several choices of scale factor, including exponential and power-law expansion, and identified conditions under which the energy conditions hold during the evolution. Kim \cite{Kim1996} embedded a wormhole in a Friedmann--Lemaitre--Robertson--Walker background by splitting the stress-energy tensor into a time-dependent cosmological part and a space-dependent wormhole part, and Anchordoqui, Torres, and Trobo \cite{AnchordoquiTorresTrobo1998} extended this construction to a broader class of evolving geometries; a recent review by Kord Zangeneh and Lobo \cite{KordZangenehLobo2025} surveys this body of work across general relativity and modified gravity. In these scale-factor constructions the shape function itself is typically static or nearly so, and the entire time dependence of the metric is carried by the conformal factor $a(t)^2$ multiplying the spatial line element; whether the energy conditions are satisfied is then controlled by the rate of expansion of the background. The construction adopted in the present work differs in that the shape function is itself the dynamical quantity, with $\beta_t(r,t)$ fixed directly by the off-diagonal Einstein equation rather than by a prescribed $a(t)$. The evolution here is therefore sourced locally, through a radial flux concentrated near the throat, instead of being driven by the global expansion of a cosmological background. The two mechanisms are not mutually exclusive: a scale-factor construction could in principle be superposed on the flux-driven shape function used here, though this introduces additional off-diagonal and time-time field equations that lie outside the scope of the present analysis.

In the present work, we adopt a different viewpoint by constructing a dynamical wormhole model in which the evolution of the geometry is driven by a radial energy flux. Instead of prescribing the time dependence of the shape function, we use the off-diagonal field equation to relate its temporal variation directly to the energy flux. A similar idea of matter-driven dynamical wormhole generation has been explored in the context of charged fluids \cite{LoboOlmoRubieraGarcia2014}.

By considering a decaying flux profile, we obtain an exact class of solutions in which the shape function consists of a static component and a transient contribution that vanishes exponentially with time. The resulting geometry satisfies the essential conditions for a traversable wormhole, including the existence of a throat, the flaring-out condition, and asymptotic flatness. We analyze the corresponding matter variables, examine the energy conditions, and quantify the amount of exotic matter through a volume integral measure. Furthermore, we investigate the dynamical stability of the configuration under linear perturbations. This approach provides a physically transparent description of dynamical wormholes as flux-driven systems that evolve toward equilibrium, and offers a simple framework for constructing time-dependent solutions within general relativity.

The paper is organized as follows. In Sec.~\ref{sec2}, we present the field equations corresponding to the dynamical wormhole spacetime and establish the relation between the temporal evolution of the geometry and the radial energy flux. In Sec.~\ref{sec3}, we examine the late-time behavior of the obtained solutions and show how the geometry approaches a static configuration. Sec.~\ref{sec4} is devoted to the quantification of exotic matter through a suitable volume integral, with particular emphasis on its time dependence and decay during the evolution. Finally, in Sec.~\ref{sec5}, we summarize the main results and discuss their physical implications.

\section{Field equations for the dynamical wormhole spacetime}\label{sec2}

We begin with the Einstein--Hilbert action in geometrized units $(G=c=1)$,
\begin{equation}
S=\int d^4x\,\sqrt{-g}\left(\frac{R}{16\pi}+\mathcal{L}_{m}\right),
\end{equation}
where $R$ is the Ricci scalar and $\mathcal{L}_{m}$ denotes the matter Lagrangian. Variation of the action with respect to the metric tensor $g_{\mu\nu}$ yields the Einstein field equations
\begin{equation}
G_{\mu\nu}=8\pi T_{\mu\nu}.
\end{equation}

To investigate time-dependent wormhole geometries, we consider a general spherically symmetric spacetime with a time-dependent shape function. To describe a time-dependent wormhole geometry, we consider the metric
\begin{equation}
ds^{2}=-e^{2\Phi(r,t)}dt^{2}+\frac{dr^{2}}{1-\beta(r,t)/r}+r^{2}\left(d\theta^{2}+\sin^{2}\theta\,d\phi^{2}\right),
\label{metric_dynamic_wh}
\end{equation}
where $\Phi(r,t)$ and $\beta(r,t)$ represent the redshift and shape functions, respectively. Introducing
\begin{equation}
e^{-b(r,t)}=1-\frac{\beta(r,t)}{r}, \qquad a(r,t)=2\Phi(r,t),
\end{equation}
the line element \eqref{metric_dynamic_wh} takes the standard spherically symmetric form
\begin{equation}
ds^{2}=-e^{a(r,t)}dt^{2}+e^{b(r,t)}dr^{2}+r^{2}d\Omega^{2}.
\end{equation}

For the anisotropic dynamical matter distribution
\begin{eqnarray}
T^{\mu}_{\ \nu}&=&
\mathrm{diag}\left[-\rho(r,t),\,p_{r}(r,t),\,p_{t}(r,t),\,p_{t}(r,t)\right]\nonumber\\
&+&\text{off-diagonal flux terms},
\end{eqnarray}
the non-vanishing Einstein equations for the above metric are
\begin{equation}
8\pi \rho=\frac{e^{-b}b'}{r}+\frac{1-e^{-b}}{r^{2}},
\label{rho_gen}
\end{equation}
\begin{equation}
8\pi p_{r}=\frac{e^{-b}a'}{r}-\frac{1-e^{-b}}{r^{2}},
\label{pr_gen}
\end{equation}
\begin{eqnarray}
8\pi p_{t}
&=&\frac{e^{-b}}{4}\left(2a''+a'^{2}-a'b'+\frac{2(a'-b')}{r}\right)\nonumber\\
&+&\frac{e^{-a}}{4}\left(\dot{a}\dot{b}-\dot{b}^{2}-2\ddot{b}\right),
\label{pt_gen}
\end{eqnarray}
while the off-diagonal component reads
\begin{equation}
8\pi T^{r}_{\ t}=\frac{e^{-b}\dot{b}}{r}.
\label{flux_gen}
\end{equation}
Here, a prime and an overdot denote differentiation with respect to $r$ and $t$, respectively.

Using
\begin{equation}
a'=2\Phi_{r}, \qquad a''=2\Phi_{rr}, \qquad \dot{a}=2\Phi_{t},
\end{equation}
together with
\begin{equation}
b'=\frac{r\beta_{r}-\beta}{r(r-\beta)}, \;
\dot{b}=\frac{\beta_{t}}{r-\beta}, \;
\ddot{b}=\frac{\beta_{tt}}{r-\beta}+\frac{\beta_{t}^{2}}{(r-\beta)^{2}},
\end{equation}

Substituting these relations into Eqs. \eqref{rho_gen}--\eqref{flux_gen} and simplifying, we obtain the following compact form of the field equations:
\begin{eqnarray}
\rho(r,t)&=&\frac{\beta_{r}(r,t)}{8\pi r^{2}},
\label{rho_dynamic}\\
p_{r}(r,t)&=&\frac{1}{8\pi}\left[\frac{2}{r}\Big[1-\frac{\beta(r,t)}{r}\Big]\Phi_{r}(r,t)-\frac{\beta(r,t)}{r^{3}}\right],\label{pr_dynamic}\\
p_{t}(r,t)&=&\frac{1}{8\pi}\Big[\Big[1-\frac{\beta(r,t)}{r}\Big]
\Big[\Phi_{rr}(r,t)+\Phi_{r}^{2}(r,t)\nonumber\\&+&\frac{\Phi_{r}(r,t)}{r}\Big]-\frac{r\beta_{r}(r,t)-\beta(r,t)}{2r^{2}}
\Big[\Phi_{r}(r,t)\nonumber\\&+&\frac{1}{r}\Big]+e^{-2\Phi(r,t)}
\Big[
\frac{\Phi_{t}(r,t)\beta_{t}(r,t)-\beta_{tt}(r,t)}{2[r-\beta(r,t)]}
\nonumber\\&-&\frac{3\beta_{t}^{2}(r,t)}{4[r-\beta(r,t)]^{2}}
\Big]
\Big],\label{pt_dynamic}\\
T^{r}_{\ t}(r,t)&=&\frac{\beta_{t}(r,t)}{8\pi r^{2}}.
\label{flux_dynamic}
\end{eqnarray}

Equations \eqref{rho_dynamic}--\eqref{flux_dynamic} represent the Einstein field equations for the dynamical wormhole metric \eqref{metric_dynamic_wh}. In the static limit, $\Phi_{t}=0$, $\beta_{t}=0$, and $\beta_{tt}=0$, the flux term vanishes and the above system reduces to the standard Morris--Thorne wormhole equations. We consider the dynamical wormhole metric and assume a constant redshift function,
\begin{equation}
\Phi(r,t)=\Phi_{0}=\text{constant}.
\end{equation}
Without loss of generality, one may set $e^{2\Phi_{0}}=1$ by rescaling the time coordinate. Under this assumption, the Einstein field equations reduce to the following set of independent equations:

\begin{eqnarray}
\rho(r,t)&=&\frac{\beta_{r}(r,t)}{8\pi r^{2}},\label{18}\\
p_{r}(r,t)&=&-\frac{\beta(r,t)}{8\pi r^{3}},\label{19}\\
p_{t}(r,t)&=&\frac{1}{8\pi}\Big[\frac{\beta(r,t)-r\,\beta_{r}(r,t)}{2r^{3}}
-\frac{\beta_{tt}(r,t)}{2\left(r-\beta(r,t)\right)}
\nonumber\\&-&\frac{3\,\beta_{t}^{2}(r,t)}{4\left(r-\beta(r,t)\right)^{2}}
\Big],\label{20}\\
T^{r}_{\ t}(r,t)&=&\frac{\beta_{t}(r,t)}{8\pi r^{2}},\label{21}
\end{eqnarray}

It is important to examine the behavior of the tangential pressure near the wormhole throat, where $r \to r_{\rm th}(t)$ and $r - \beta(r,t) \to 0$. At first sight, the expression for $p_t$ contains terms proportional to $(r-\beta)^{-1}$ and $(r-\beta)^{-2}$, which may suggest a divergence.

However, a careful limiting analysis shows that these apparent singularities are canceled by corresponding factors in the numerator. Expanding the shape function near the throat as
\begin{equation}
\beta(r,t)=r_{\rm th}(t) + \beta_r(r_{\rm th},t)(r-r_{\rm th}) + \cdots,
\end{equation}
we find that the combination $r-\beta(r,t)$ behaves linearly in $(r-r_{\rm th})$.

Substituting this expansion into the expression for $p_t$, one observes that the leading divergent terms cancel, leaving a finite result at the throat. Therefore, the tangential pressure remains regular provided that the flaring-out condition $\beta_r(r_{\rm th},t)<1$ is satisfied.

This ensures that the matter variables are well-behaved and the wormhole geometry is free from physical singularities at the throat.

Equations~\eqref{18}--\eqref{21} represent the field equations governing a dynamical wormhole supported by anisotropic matter with radial energy flux in the zero-tidal-force case. In the static limit, $\beta_{t}=0$ and $\beta_{tt}=0$, the flux term vanishes and the above system reduces to the standard Morris--Thorne wormhole equations.

In a dynamical wormhole spacetime, the time dependence of the geometry implies that the matter distribution cannot remain static. In particular, any variation of the shape function with respect to time must be accompanied by a flow of energy along the radial direction. This is reflected in the off-diagonal component of the energy-momentum tensor, $T^r_{\ t}$, which represents the radial energy flux. Therefore, a non-vanishing flux term is not an arbitrary assumption, but a direct consequence of considering a time-dependent wormhole configuration. Physically, this flux can be interpreted as a mechanism through which energy is redistributed within the spacetime, driving the evolution of the geometry and allowing the system to relax toward a static equilibrium state.

An alternative method to determine the dynamical shape function can be obtained from the off-diagonal field equation. In particular, Eq.~(\ref{21}) yields
\begin{equation}
\beta_t(r,t)=8\pi r^2\,T^r_{\ t}(r,t),
\label{beta_flux_eq}
\end{equation}
which directly relates the temporal evolution of the shape function to the radial energy flux. If the flux profile $T^r_{\ t}(r,t)$ is prescribed, Eq.~\eqref{beta_flux_eq} can be integrated with respect to time to obtain
\begin{equation}
\beta(r,t)=8\pi r^2 \int T^r_{\ t}(r,t)\,dt + g(r),
\label{beta_flux_solution}
\end{equation}
where $g(r)$ is an arbitrary function of the radial coordinate determined from suitable boundary or matching conditions. This formulation provides a physically transparent interpretation in which the dynamical behavior of the wormhole geometry is governed by the flow of energy along the radial direction. As an illustrative example, consider a separable flux profile of the form
\begin{equation}
T^r_{\ t}(r,t)=\frac{h(r)}{8\pi r^2}e^{-\gamma t},
\end{equation}
where $h(r)$ is an arbitrary function and $\gamma>0$ is a constant parameter controlling the temporal decay. This choice represents a decaying radial energy flux, which models a transient energy transfer process. The exponential factor ensures that the flux is significant at early times and gradually diminishes, allowing the system to approach equilibrium. Substituting this into Eq.~\eqref{beta_flux_eq}, one obtains
\begin{equation}
\beta_t(r,t)=h(r)e^{-\gamma t}.
\end{equation}
Integrating with respect to $t$, the corresponding shape function is given by
\begin{equation}
\beta(r,t)=-\frac{h(r)}{\gamma}e^{-\gamma t}+g(r).
\end{equation}

To construct a physically consistent dynamical wormhole model within the flux-driven framework, it is convenient to specify explicit functional forms for the radial functions $g(r)$ and $h(r)$ appearing in the general solution. Here, $g(r)$ represents the static component of the geometry, while $h(r)$ governs the radial profile of the dynamical contribution induced by the energy flux.  A simple yet physically meaningful choice is to take
\begin{equation}
g(r)=r_0\left(\frac{r_0}{r}\right)^{n}, \qquad n>0,
\label{g_choice}
\end{equation}
which corresponds to a well-behaved static wormhole shape function. This form satisfies the essential geometric requirements, namely
\begin{equation}
g(r_0)=r_0, \qquad g'(r_0)=-\frac{n}{r_0}<1,
\end{equation}
and ensures asymptotic flatness since $g(r)/r \to 0$ as $r \to \infty$. For the dynamical part, we choose
\begin{equation}
h(r)=\eta\,r_0\left(\frac{r_0}{r}\right)^{m}, \qquad m>0,
\label{h_choice}
\end{equation}
where $\eta$ is a dimensionless parameter controlling the strength of the dynamical perturbation. This form guarantees that the time-dependent correction remains finite near the throat and decays at large radial distances. Substituting Eqs.~\eqref{g_choice} and \eqref{h_choice} into the general solution, the dynamical shape function takes the form
\begin{equation}
\beta(r,t)=r_0\left(\frac{r_0}{r}\right)^{n}
-\frac{\eta}{\gamma}\,r_0\left(\frac{r_0}{r}\right)^{m}e^{-\gamma t}.
\label{beta_model}
\end{equation}

The negative sign of the time-dependent term requires some care. At early times this term lowers the value of the shape function and may shift the throat away from the reference scale $r_0$. For some parameter choices, it may even remove the real positive root of $\beta(r,t)=r$. Therefore, the parameters $\eta$, $\gamma$, $m$, and $n$ must be chosen so that the static part of the geometry is not overwhelmed by the transient correction near the throat. A simple sufficient restriction follows from the value of the shape function near $r=r_0$,
\begin{equation}
1-\frac{\eta}{\gamma}e^{-\gamma t}>0.
\end{equation}
This condition shows that the ratio $\eta/\gamma$ should remain sufficiently small, especially at early times. Physically, this means that the radial flux should act as a controlled dynamical correction rather than destroying the underlying wormhole geometry.

The parameters $n$ and $m$ determine how rapidly the static and time-dependent parts of the shape function decrease with the radial coordinate, whereas $\eta$ fixes the magnitude of the dynamical contribution. The term $g(r)$ defines the static background geometry, while $h(r)$ gives the radial profile of the flux-induced correction. For $\gamma>0$, the exponential factor causes the dynamical part to decay with time, and hence
\begin{equation}
\lim_{t \to \infty}\beta(r,t)=g(r).
\end{equation}
Thus, the solution describes a transient wormhole geometry that approaches a static configuration at late times.

For the dynamical model, the throat cannot in general be identified with $r_0$ at finite time. Instead, the instantaneous throat radius is determined by
\begin{equation}
\beta(r_{\rm th}(t),t)=r_{\rm th}(t).
\label{dyn_throat_condition}
\end{equation}
Therefore, $r_{\rm th}$ is a time-dependent quantity. The parameter $r_0$ should be understood as the reference scale of the static background and becomes the throat radius only in the late-time limit. For the parameter values used in the numerical analysis, Eq.~\eqref{dyn_throat_condition} admits a real and positive solution during the evolution, confirming that the wormhole throat remains well defined.

\begin{figure}[!htbp]
\includegraphics[width=7.5cm, height=5.5cm]{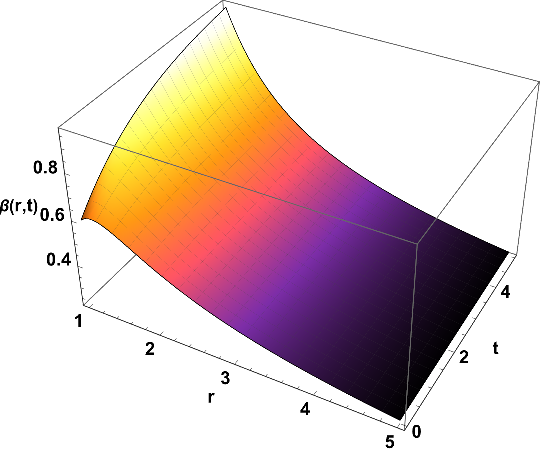}
\includegraphics[width=7.5cm, height=5.5cm]{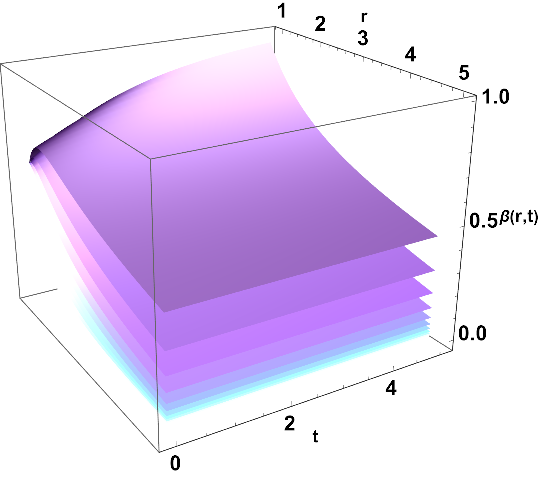}
\includegraphics[width=1cm, height=5.5cm]{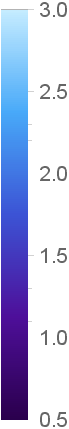}
\includegraphics[width=7.5cm, height=5.5cm]{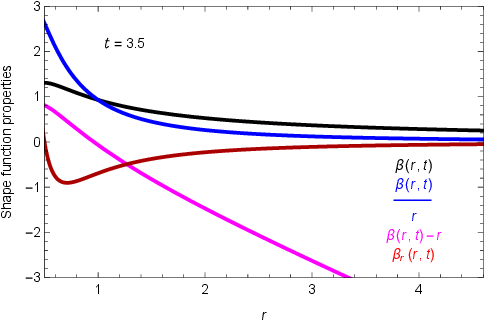}
\caption{Plots of the wormhole shape function upper graph for fixed $n$ along radial coordinate $r$ and time $t$, wormhole shape function middle graph for different values of $n$ along radial coordinate $r$ and time $t$, properties of shape function lower graph versus $r$ for fixed $t$ by fixing the parameters $\gamma =0.5\;\eta =0.2\;m=3\;n=0.9\;r_0=1$.}\label{Fig:1}
\end{figure}

Fig. \ref{Fig:1} illustrates the behavior of the proposed dynamical shape function and the associated geometric conditions. The upper panel shows the variation of $\beta(r,t)$ with both $r$ and $t$ for fixed parameter values. The surface indicates that the time-dependent correction is stronger at early times and gradually decreases as $t$ increases. This confirms the relaxation of the dynamical geometry toward its static background form.

The middle panel displays the behavior of the shape function for different values of the parameter $n$. Increasing $n$ changes the radial fall-off of the static component and therefore modifies how sharply the shape function decreases away from the throat region. This shows that $n$ controls the radial compactness of the static wormhole geometry, whereas the parameters $\eta$ and $\gamma$ determine the size and decay rate of the transient deformation.

The lower panel of Fig.~\ref{Fig:1} presents the main wormhole conditions. The curve corresponding to $\beta(r,t)-r$ is used to locate the instantaneous throat through the root of Eq.~\eqref{dyn_throat_condition}. Since the root is not fixed at $r=r_0$ for finite $t$, the figure should be interpreted as showing the dynamical position of the throat rather than a permanently fixed throat radius. The ratio $\beta(r,t)/r$ decreases with increasing $r$ and tends toward zero in the asymptotic region, which supports the asymptotic flatness of the geometry. The derivative $\beta_r(r,t)$ remains below unity near the throat for the chosen parameters, indicating that the flaring-out condition is satisfied. In addition, the condition $\beta(r,t)<r$ outside the throat is maintained, so no horizon appears in the plotted region. These features show that the proposed parameter set lies within the physically acceptable domain for a dynamical traversable wormhole.

\begin{figure}[!htbp]
\includegraphics[width=7.5cm, height=5.5cm]{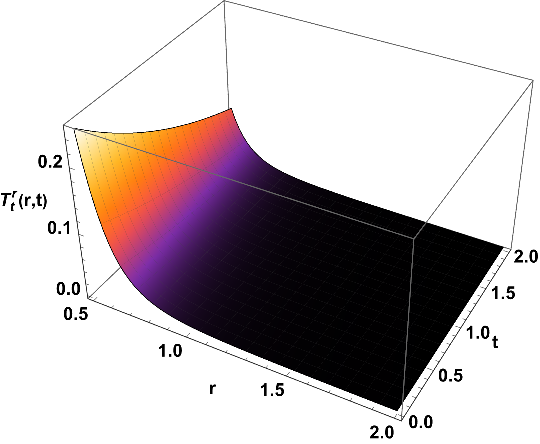}
\caption{Profiles of flux along radial coordinate $r$ and time $t$ by fixing the parameters $\gamma =0.5\;\eta =0.2\;m=3\;n=0.9\;r_0=1$.}\label{Fig:2}
\end{figure}

\begin{figure}[!htbp]
\includegraphics[width=7.5cm, height=5.5cm]{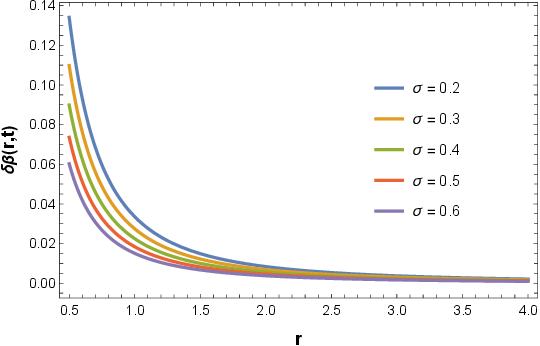}
\includegraphics[width=7.5cm, height=5.5cm]{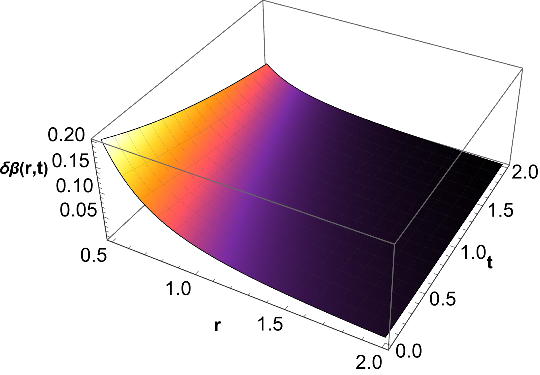}
\caption{Perturbation profiles upper graph for different $\sigma$ along radial coordinate $r$, and lower graph for fixed $\sigma$ along radial coordinate $r$ and time $t$ by fixing the parameters $s=2\;\delta_0=0.5\;r_0=1s$.}\label{Fig:3}
\end{figure}

\begin{figure}[!htbp]
\includegraphics[width=7.5cm, height=5.5cm]{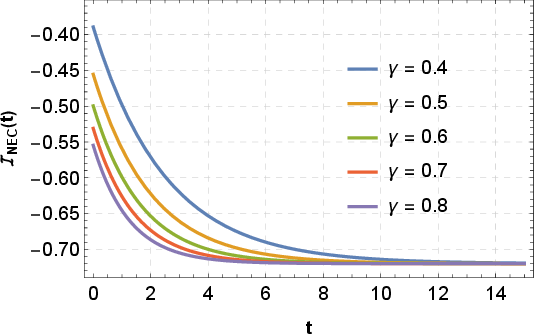}
\caption{Integral quantifier profiles along radial coordinate $r$ and time $t$ by fixing the parameters $\gamma =0.5\;\eta =0.2\;m=3\;n=0.9\;r_0=1$.}\label{Fig:4}
\end{figure}

\begin{figure}[!htbp]
\includegraphics[width=7.5cm, height=5.5cm]{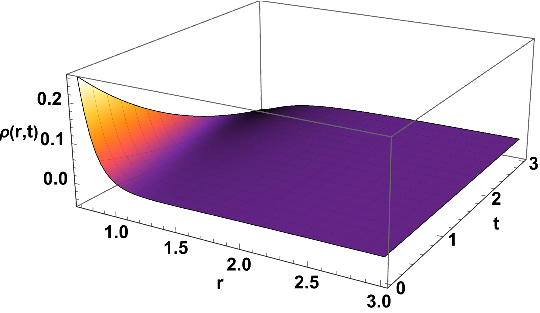}
\includegraphics[width=7.5cm, height=5.5cm]{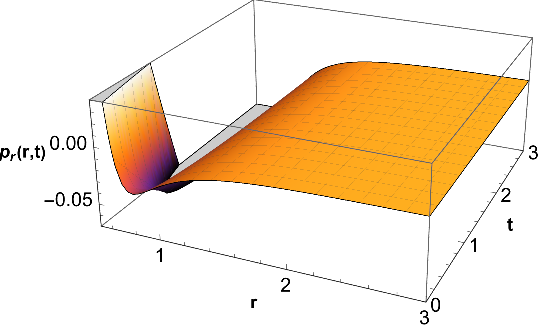}
\includegraphics[width=7.5cm, height=5.5cm]{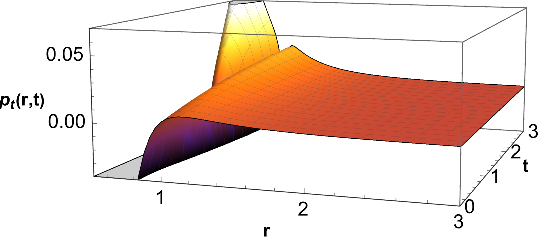}
\caption{Energy conditions profiles along radial coordinate $r$ and time $t$ by fixing the parameters $\gamma =0.5\;\eta =0.2\;m=3\;n=0.9\;r_0=1$.}\label{Fig:5}
\end{figure}

\begin{figure}[!htbp]
\includegraphics[width=7.5cm, height=5.5cm]{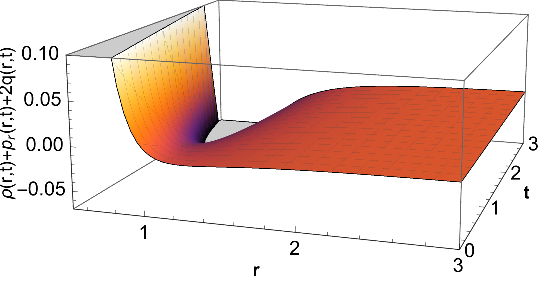}
\includegraphics[width=7.5cm, height=5.5cm]{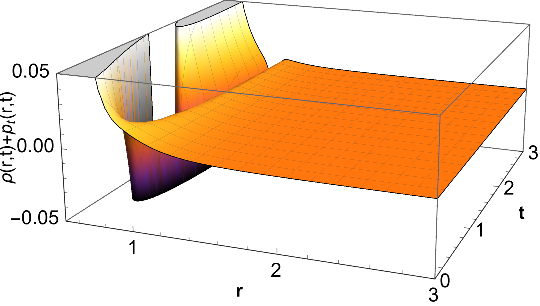}
\includegraphics[width=7.5cm, height=5.5cm]{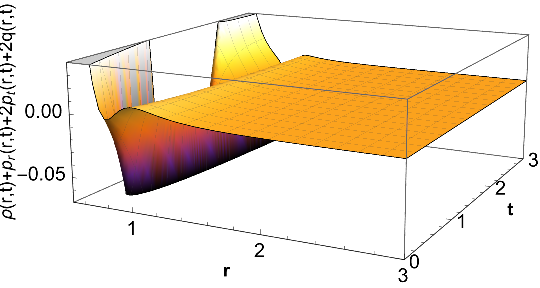}
\caption{Energy conditions profiles along radial coordinate $r$ and time $t$ by fixing the parameters $\gamma =0.5\;\eta =0.2\;m=3\;n=0.9\;r_0=1$.}\label{Fig:6}
\end{figure}

Replacing Eq. (\ref{beta_model}) in set of Eqs. \eqref{rho_dynamic}-\eqref{flux_dynamic}, we get the final version of equations as:
\begin{eqnarray}
    \rho(r,t)&=&-\frac{\gamma  n r_0 \left(\frac{r_0}{r}\right)^n-\eta  m r_0 e^{\gamma  (-t)} \left(\frac{r_0}{r}\right)^m}{8 \pi  \gamma  r^3},\\
    p_r(r,t)&=& -\frac{r_0 \left(\left(\frac{r_0}{r}\right)^n-\frac{\eta  e^{\gamma  (-t)} \left(\frac{r_0}{r}\right)^m}{\gamma }\right)}{8 \pi  r^3},\\
    p_t(r,t)&=& \frac{r_0}{32 \pi } \Big[\frac{2}{r^3} \Big[(n+1) \left(\frac{r_0}{r}\right)^n-\frac{\eta}{\gamma }  (m+1) \nonumber\\&\times&e^{\gamma  (-t)} \left(\frac{r_0}{r}\right)^m\Big]-\frac{1}{f(r,t)}\Big[\gamma ^2 \eta  \left(\frac{r_0}{r}\right)^m\nonumber\\&\times& \Big[\eta  r_0 \left(\frac{r_0}{r}\right)^m-2 \gamma  e^{\gamma  t} \Big[r-r_0 \left(\frac{r_0}{r}\right)^n\Big]\Big]\Big]\Big] ,\\
    T^r_t(r,t)&=&\frac{\eta r_0}{8 \pi  r^2} e^{\gamma  (-t)} \left(\frac{r_0}{r}\right)^m,
\end{eqnarray}
where
\begin{eqnarray*}
    f(r,t)=\left(\eta  r_0 \left(\frac{r_0}{r}\right)^m+\gamma  e^{\gamma  t} \left(r-r_0 \left(\frac{r_0}{r}\right)^n\right)\right)^2.
\end{eqnarray*}

Fig. \ref{Fig:2} illustrates the radial energy flux, which is seen to be strongest near the throat and decreases both with radial distance and time. This behavior confirms that the dynamical evolution is driven by a localized energy transport mechanism.

\subsection{Possible physical origin of the radial flux}

The flux $T^r_{\ t}(r,t)$ used above is reconstructed from a prescribed shape function rather than derived from a specified matter Lagrangian, and it is worth indicating what type of source could generate a flux of this form. A radial, time-decaying energy flow of the kind adopted here arises naturally in null-fluid models, in which energy is carried by ingoing or outgoing radiation rather than by matter at rest. Hayward \cite{Hayward2002} constructed static wormholes supported by counter-propagating beams of negative-energy null radiation and showed that switching the radiation off causes the geometry to collapse to a Schwarzschild black hole; the reverse process, in which such radiation is switched on and subsequently decays, is close in spirit to the exponentially decaying flux of Eq.~(25). A fully dynamical construction along these lines was carried out by Lobo, Olmo, and Rubiera-Garcia \cite{LoboOlmoRubieraGarcia2014}, who generated a wormhole from a charged null fluid within an extension of general relativity: while the flux is active the metric passes through a Vaidya-type region, and once the flux switches off the geometry settles into a static wormhole. The present model can be regarded as an effective, geometry-first description of a process of this kind. Rather than integrating a specific null-fluid or charged-fluid source through the field equations, we prescribe the resulting flux profile directly and solve for the corresponding shape function; a first-principles derivation of $T^r_{\ t}(r,t)$ from an explicit matter Lagrangian, along the lines of these null-fluid constructions, is left for future work.

\subsection{Dynamical stability under linear perturbations}

To examine the dynamical stability of the flux-driven wormhole configuration, we introduce a small perturbation in the shape function as
\begin{equation}
\beta(r,t)=\beta_0(r,t)+\epsilon\,\delta\beta(r,t),
\qquad 0<\epsilon\ll 1,
\end{equation}
where $\beta_0(r,t)$ represents the background solution and $\delta\beta(r,t)$ denotes a linear perturbation. This decomposition allows us to analyze how the wormhole geometry responds to small deviations from equilibrium.

The evolution of the system is governed by the off-diagonal field equation
\begin{equation}
T^r_{\ t}=\frac{\beta_t}{8\pi r^2},
\end{equation}
which directly links the time variation of the geometry to the radial energy flux. Substituting the perturbed form into this equation and retaining only first-order terms in $\epsilon$, we obtain
\begin{equation}
\delta\beta_t=8\pi r^2\,\delta T^r_{\ t}.
\end{equation}
This relation shows that the dynamics of the perturbation is controlled by the variation of the radial energy flux.

At this stage, it is important to relate the perturbation of the flux to the underlying matter dynamics. The stress-energy tensor satisfies the conservation equation $\nabla_\mu T^{\mu}{}_{\nu}=0$, which governs the evolution of the energy flux in response to changes in the geometry. In the absence of a detailed microphysical model, we adopt a linear response description in which small deviations from equilibrium induce a restoring flux proportional to the perturbation itself. This provides an effective way to capture the dissipative behavior of the matter distribution.

To model the response of the matter distribution, we assume that the perturbed flux has a restoring character, such that deviations from equilibrium are suppressed. Within this framework, this leads to the effective relation
\begin{equation}
\delta T^r_{\ t}=-\frac{\sigma}{8\pi r^2}\,\delta\beta,
\qquad \sigma>0,
\end{equation}
where $\sigma$ is a constant damping parameter that characterizes the relaxation rate of the system. This relation is analogous to a dissipative response in which the energy transport acts to reduce geometric distortions. Substituting this expression into the linearized field equation
\begin{equation}
\delta\beta_t = 8\pi r^2\,\delta T^r{}_{t},
\end{equation}
leads to
\begin{equation}
\delta\beta_t + \sigma\,\delta\beta = 0,
\end{equation}
with the solution
\begin{equation}
\delta\beta(r,t)=\delta\beta(r,0)e^{-\sigma t}.
\end{equation}

Thus, the exponential decay of perturbations follows from a simple and physically motivated relaxation model for the flux. The stability of the wormhole configuration is therefore associated with the ability of the matter distribution to dissipate perturbations through radial energy transport.

It should be stressed that Eqs.~(42) and (43) are not derived from the Einstein field equations or from a specific microphysical matter model: the relaxation law relating $\delta T^r_{\ t}$ to $\delta\beta$ is introduced as a closure assumption, in the spirit of an Eckart-type relativistic dissipative flux \cite{Eckart1940}, in which a thermodynamic flux is taken proportional to the instantaneous deviation from equilibrium rather than obtained from an independent evolution equation for the flux itself. First-order closures of this kind are known to admit instabilities and acausal signal propagation when treated as exact evolution equations rather than as a low-frequency effective description \cite{Israel1976,IsraelStewart1979}; a fully causal treatment would promote $\delta T^r_{\ t}$ to an independent dynamical variable obeying its own relaxation equation, as in Israel-Stewart theory, rather than slaving it algebraically to $\delta\beta$. The analysis presented here should accordingly be read as an effective, phenomenological stability argument, applicable on time scales longer than any microscopic relaxation time associated with the underlying flux, and not as a rigorous linear perturbation analysis of the coupled Einstein-matter system. Within this restricted sense, the result establishes that a restoring flux of the assumed form is sufficient to damp shape-function perturbations; it does not establish that such a restoring flux is necessary, nor does it exclude oscillatory or growing modes that a causal second-order treatment might reveal.

At this stage, in order to obtain explicit profiles suitable for visualization, we specify the initial radial dependence of the perturbation. We take
\begin{equation}
\delta\beta(r,0)=\delta_0\left(\frac{r_0}{r}\right)^s,
\qquad \delta_0\ll 1,\quad s>0,
\end{equation}
which ensures that the perturbation remains finite at the throat and decays at large radial distances, in agreement with the asymptotic behavior of the background geometry. The parameter $s$ controls the radial localization of the disturbance, with larger values corresponding to perturbations that are more strongly concentrated near the throat.

Substituting this profile into the general solution yields
\begin{equation}
\delta\beta(r,t)=\delta_0\left(\frac{r_0}{r}\right)^s e^{-\sigma t},
\end{equation}
which is the expression used in the numerical analysis and graphical illustrations.

The above result shows that, for $\sigma>0$, the perturbation decays exponentially with time, indicating that the wormhole configuration is dynamically stable under small disturbances. In contrast, $\sigma<0$ leads to an exponential growth of $\delta\beta(r,t)$, signaling instability.

From a physical perspective, the quantity $\delta\beta(r,t)$ represents a small distortion of the wormhole geometry. Through the field equations, such a deformation modifies the associated matter distribution. In particular, since $T^r_{\ t}$ is directly related to $\beta_t$, the perturbation corresponds to a fluctuation in the radial energy flux. The exponential decay of $\delta\beta(r,t)$ therefore reflects a dissipative process in which energy transport acts to restore the geometry toward equilibrium.

Thus, the damping behavior is not introduced in an ad hoc manner but arises as an effective description of a relaxation process in the matter sector. A more detailed determination of the parameter $\sigma$ would require specifying the microphysical properties of the matter source, which is beyond the scope of the present analysis.

It is instructive to separate the influence of the parameters $\gamma$ and $\sigma$. The parameter $\gamma$ controls how the underlying background configuration relaxes in time, whereas $\sigma$ determines the rate at which small perturbations decay. Together, they govern both the approach to equilibrium and the stability of the system.

The time evolution of the perturbation $\delta\beta(r,t)$ for various choices of the parameter $\sigma$ is presented in Fig.~\ref{Fig:3}. It is observed that larger positive values of $\sigma$ correspond to a more rapid suppression of the perturbation, while negative values lead to an amplification over time, consistent with the analytical behavior obtained earlier. This confirms that the configuration remains stable when $\sigma>0$, as small deviations are progressively damped. Furthermore, the spatial profiles show that the perturbation is localized near the throat and decreases with increasing radial distance. The exponential time dependence clearly demonstrates the dissipative nature of the evolution.

\section{Late-time behavior}\label{sec3}
To understand the final state of the dynamical configuration, we examine the late-time limit $t \to \infty$. In this limit, the exponential factor $e^{-\gamma t}$ vanishes, and the shape function reduces to
\begin{equation}
\beta(r,t) \longrightarrow r_0\left(\frac{r_0}{r}\right)^n.
\end{equation}
This shows that the wormhole geometry approaches a static configuration at late times.

From the field equation $T^r_{\ t}=\beta_t/(8\pi r^2)$, it follows that the radial energy flux also vanishes in this limit, indicating that the system ceases to exchange energy. Consequently, the matter variables become time-independent and the spacetime settles into equilibrium.

Physically, this behavior reflects a relaxation process in which the dynamical contributions decay over time, leaving behind a stable wormhole supported by a static matter distribution. The parameter $\gamma$ controls the rate at which this relaxation occurs. This result shows that the dynamical contribution acts as a transient deformation, which vanishes at late times, leaving a stable wormhole geometry.

\section{Exotic matter quantification and its time decay}\label{sec4}
To quantify the amount of exotic matter supporting the dynamical wormhole, we introduce the volume integral quantifier
\begin{equation}
\mathcal{I}(t)=\int_{r_0}^{R}\left[\rho(r,t)+p_r(r,t)\right]4\pi r^2\,dr .
\end{equation}
Unlike the local NEC condition, this quantity measures the integrated contribution of NEC-violating matter in the region surrounding the throat. Since the dynamical sector of the present model is exponentially suppressed, the integral can be decomposed into a static contribution and a transient part proportional to \(e^{-\gamma t}\). Therefore, the model allows one to track how the amount of exotic matter evolves during the relaxation process.

The evolution of the volume integral quantifier is shown in Fig. \ref{Fig:4}. It is observed that
\(\mathcal{I}_{\rm NEC}(t)\) approaches a constant value at late times, which confirms that the transient part of the exotic matter contribution decays exponentially. Larger values of \(\gamma\) lead to faster relaxation, indicating that the parameter \(\gamma\) controls the rate at which the system approaches its static configuration. This provides a quantitative measure of how the amount of exotic matter evolves during the dynamical process.

\subsection{Energy conditions in the dynamical anisotropic wormhole}

An important consistency requirement for any physically acceptable matter source in general relativity is the conservation of the stress-energy tensor,
\begin{equation}
\nabla_\mu T^{\mu\nu}=0.
\end{equation}
In the present model, the matter variables $\rho(r,t)$, $p_r(r,t)$, $p_t(r,t)$, and the flux component $T^r_{\ t}(r,t)$ are not introduced independently, but are reconstructed directly from the Einstein field equations.

As a result, the conservation equations are automatically satisfied through the contracted Bianchi identities. Nevertheless, it is useful to verify that the obtained configuration is physically consistent. In particular, the radial component of the conservation equation relates the time variation of the flux to the gradients of the pressure and the geometry, ensuring that the matter distribution behaves as an anisotropic fluid with radial energy transport. This confirms that the reconstructed stress-energy tensor provides a self-consistent description of the matter supporting the dynamical wormhole.

To assess the physical properties of the matter supporting the wormhole, we analyze the standard energy conditions. In the present case, the spacetime is explicitly dynamical and contains a non-vanishing radial energy flux, which modifies the usual interpretation of these conditions.

In particular, the stress-energy tensor is not purely diagonal but includes an off-diagonal component associated with radial energy transport. It can be written as
\begin{equation}
T^{\mu}_{\ \nu}=
\begin{pmatrix}
-\rho(r,t) & 0 & 0 & 0 \\
T^r_{\ t}(r,t) & p_r(r,t) & 0 & 0 \\
0 & 0 & p_t(r,t) & 0 \\
0 & 0 & 0 & p_t(r,t)
\end{pmatrix},
\end{equation}
where the quantity
\begin{equation}
q(r,t)=T^r_{\ t}(r,t)=\frac{\beta_t(r,t)}{8\pi r^2}
\end{equation}
represents the radial energy flux. This flux depends on both the radial coordinate and time, and therefore plays a central role in the dynamical evolution of the system.

The energy density $\rho$, radial pressure $p_r$, and tangential pressure $p_t$ are obtained from the field equations. The null energy condition (NEC) requires that $T_{\mu\nu}k^\mu k^\nu \geq 0$ for any null vector $k^\mu$.

For a radial null vector chosen along the outward direction, the NEC takes the form
\begin{equation}
\rho + p_r + 2 q(r,t) \geq 0.
\end{equation}
This expression shows that the violation of the NEC is influenced not only by the pressure components but also by the radial energy flux.

In general, the NEC is violated near the throat, as required for the existence of a traversable wormhole. However, the magnitude of this violation decreases with time, indicating a gradual reduction in the effective exotic matter content as the dynamical contribution decays.

The weak energy condition (WEC) requires that $T_{\mu\nu}u^\mu u^\nu \geq 0$ for any timelike vector $u^\mu$. This leads to

\begin{equation}
\rho \geq 0,
\qquad
\rho + p_r + 2 q(r,t) \geq 0,
\qquad
\rho + p_t \geq 0.
\end{equation}

These conditions ensure that the energy density measured by any observer remains non-negative when the flux contribution is included. In the present configuration, the energy density $\rho(r,t)$ can become negative near the throat, leading to violation of the WEC, which is consistent with the requirement of exotic matter in wormhole physics.

The strong energy condition (SEC) is given by

\begin{equation}
\rho + p_r + 2p_t + 2 q(r,t) \geq 0,
\end{equation}

and is associated with the attractive nature of gravity. Due to the combined effect of negative radial pressure and the flux contribution, this condition is generally violated near the throat. This violation reflects the effective repulsive behavior required to prevent gravitational collapse and sustain the wormhole structure.

The dominant energy condition (DEC) requires that the energy flux does not exceed the energy density, leading to

\begin{equation}
\rho \geq |p_r + q(r,t)|,
\qquad
\rho \geq |p_t|.
\end{equation}

For the present model, these conditions are typically not satisfied, indicating that the matter supporting the wormhole is non-standard and does not correspond to ordinary classical matter.

An important feature of the present model is the explicit time dependence of all matter variables. Each quantity can be decomposed as
\begin{equation}
X(r,t)=X_{\text{static}}(r)+X_{\text{dynamic}}(r)\,e^{-\gamma t},
\end{equation}
where $X$ represents $\rho$, $p_r$, or $p_t$ as well as the flux $q(r,t)$.

As $t \to \infty$, the dynamical contributions vanish, including the radial energy flux, and the system approaches a static configuration. This implies that the violations of the energy conditions evolve with time and gradually settle to their static values.

The above results indicate that the wormhole is supported by exotic matter, as expected from the violation of the energy conditions. However, the presence of the radial energy flux introduces a time-dependent correction that reduces the magnitude of the violation during the evolution. As the system relaxes, this contribution decays, leaving behind a stable static configuration.

In the present work, the graphical analysis focuses on the diagonal combinations of the stress-energy tensor. The effect of the radial energy flux is incorporated analytically through the modified energy-condition expressions given above.

Figs. \ref{Fig:5} and \ref{Fig:6} show the profiles of the energy conditions. It is observed that the null and weak energy conditions are violated near the throat, while their magnitude decreases with increasing radial coordinate and time. This behavior reflects the localized nature of exotic matter and its gradual reduction during the dynamical evolution.

\section{Conclusion}\label{sec5}

In this study, we have explored a time-dependent wormhole configuration in general relativity by treating the shape function as a function of both the radial coordinate and time. The temporal evolution of the spacetime is controlled by the off-diagonal component of the Einstein equations, which connects the rate of change of the geometry to a radial flow of energy. In this way, the dynamical behavior of the wormhole is not imposed artificially, but arises from an underlying energy transport process that drives the system away from, and eventually back toward, equilibrium.

By specifying a suitable form for the energy flux, we derived an exact solution for the evolving shape function, consisting of a stationary part together with a time-dependent correction. The latter decreases exponentially, allowing the geometry to change smoothly before settling into a static configuration at late times. A detailed examination of the solution shows that the essential requirements for a traversable wormhole, such as the existence of a throat, the flaring-out condition, and asymptotic flatness, remain satisfied during the entire evolution.

The corresponding matter variables were derived explicitly, and their time dependence reflects the influence of the energy flux. The energy condition analysis shows that the null and weak energy conditions are violated in the vicinity of the throat, as expected for traversable wormholes. However, the magnitude of this violation is reduced during the dynamical phase due to the presence of the flux-driven term, which gradually decays with time. This behavior indicates that the model allows a transient reduction in the amount of exotic matter, while still maintaining the necessary conditions for wormhole existence.

To quantify this effect, we introduced the volume integral of the null energy condition, which measures the total amount of exotic matter in a finite region. It was shown that this quantity consists of a static component and a time-dependent contribution that decreases exponentially. As a result, the wormhole undergoes a relaxation process in which the matter distribution evolves toward a stable configuration with a finite and well-defined exotic matter content.

We further examined the dynamical stability of the solution under linear perturbations. By relating the perturbation of the shape function to fluctuations in the radial energy flux, we obtained a simple evolution equation that admits exponentially decaying solutions. This demonstrates that the wormhole configuration is dynamically stable for positive damping parameter, as small disturbances are naturally suppressed over time. \rev{As discussed above, this stability argument is phenomenological in the sense of Eckart-type dissipative closures \cite{Eckart1940,Israel1976,IsraelStewart1979} rather than derived from a specific matter model, and its range of validity should be understood accordingly.}

The overall picture that emerges is that of a self-consistent dynamical wormhole, in which the geometry, matter distribution, and energy transport are closely interconnected. The radial energy flux acts as a driving mechanism that regulates the evolution and facilitates the transition toward equilibrium. This provides a physically transparent interpretation of time-dependent wormhole solutions, going beyond purely static constructions. The present framework differs from standard treatments in that the dynamical behavior of the wormhole is not imposed, but emerges naturally from the prescribed energy flux. It also differs from the cosmological, scale-factor-based evolving wormholes reviewed in \cite{KordZangenehLobo2025}, in which the time dependence is instead carried by an overall conformal factor multiplying an otherwise static shape function; the two mechanisms of evolution, a local radial flux and a global cosmological expansion, are complementary rather than competing, and could in principle be combined.

Two aspects of the present treatment are limitations rather than results and are worth restating explicitly. First, the linear stability analysis of Sec.~\ref{sec2} rests on a phenomenological relaxation law for the perturbed flux rather than on a matter model derived from a Lagrangian, and is valid only on time scales longer than any microscopic relaxation time of the flux itself. Second, the flux $T^r_{\ t}(r,t)$ driving the evolution is reconstructed from the assumed shape function rather than obtained from a specified microphysical source; null-fluid and charged-fluid constructions \cite{Hayward2002,LoboOlmoRubieraGarcia2014} indicate a plausible physical realization along Vaidya-type lines, but a first-principles derivation from an explicit matter Lagrangian is left for future work.

The approach presented in this work suggests a number of possible extensions. For example, one may consider applying the same flux-based construction within alternative theories of gravity to examine how the underlying dynamics are modified. It would also be worthwhile to investigate different types of matter content and their influence on the evolution of the geometry. In addition, the time-dependent nature of the solutions raises the question of whether such configurations could leave observable imprints during their transient phase. Exploring these aspects may help to clarify the physical relevance of wormhole models and their potential role in gravitational phenomena.

\section*{Funding}
This work has received funding support from the NSRF via the Program Management Unit for Human Resources \& Institutional Development, Research and Innovation [grant number B39G690007].

\end{document}